\pdfoutput=1
\documentclass[aps,prl,reprint,superscriptaddress]{revtex4-1}
\usepackage{graphicx}
\usepackage{MnSymbol}%
\usepackage{wasysym}%
\usepackage{bm}
\usepackage{braket}
\usepackage{url}

\usepackage[breaklinks]{hyperref}
\hypersetup{
colorlinks=true,final=true,
        linkcolor=blue,
        citecolor=blue,
        filecolor=blue,
        urlcolor=blue,
}

\begin{document}

\title{Coherent spin-qubit photon coupling}


\author{A. J. Landig}
\thanks{These authors contributed equally to this work.}
\affiliation{Department of Physics, ETH Z\"urich, CH-8093 Z\"urich, Switzerland}
\author{J. V. Koski}
\thanks{These authors contributed equally to this work.}
\affiliation{Department of Physics, ETH Z\"urich, CH-8093 Z\"urich, Switzerland}
\author{P. Scarlino}
\affiliation{Department of Physics, ETH Z\"urich, CH-8093 Z\"urich, Switzerland}
\author{U. C. Mendes}
\affiliation{Institut quantique and D\'{e}partment de Physique, Universit\'{e} de Sherbrooke,Sherbrooke, Qu\'{e}bec J1K 2R1, Canada}
\author{A. Blais}
\affiliation{Institut quantique and D\'{e}partment de Physique, Universit\'{e} de Sherbrooke,Sherbrooke, Qu\'{e}bec J1K 2R1, Canada}
\affiliation{Canadian Institute for Advanced Research, Toronto, ON, Canada}
\author{C. Reichl}
\affiliation{Department of Physics, ETH Z\"urich, CH-8093 Z\"urich, Switzerland}
\author{W. Wegscheider}
\affiliation{Department of Physics, ETH Z\"urich, CH-8093 Z\"urich, Switzerland} 
\author{A. Wallraff}
\affiliation{Department of Physics, ETH Z\"urich, CH-8093 Z\"urich, Switzerland} 
\author{K. Ensslin}
\affiliation{Department of Physics, ETH Z\"urich, CH-8093 Z\"urich, Switzerland} 
\author{T. Ihn}
\affiliation{Department of Physics, ETH Z\"urich, CH-8093 Z\"urich, Switzerland}

%\date{\today}

\begin{abstract}
Electron spins hold great promise for quantum computation due to their long coherence times. An approach to realize interactions between distant spin-qubits is to use photons as carriers of quantum information. We demonstrate strong coupling between single microwave photons in a NbTiN high impedance cavity and a three-electron spin-qubit in a GaAs triple quantum dot. We resolve the vacuum Rabi mode splitting with a coupling strength of $g/2\pi\simeq31\,\mathrm{MHz}$ and a qubit decoherence of $\gamma_2/2\pi\simeq 20\,\mathrm{MHz}$. We can tune the decoherence electrostatically and obtain a minimal $\gamma_2/2\pi\simeq 10\,\mathrm{MHz}$ for $g/2\pi\simeq 23\,\mathrm{MHz}$. The dependence of the qubit-photon coupling strength on the tunable electric dipole moment of the qubit is measured directly using the ac Stark effect. Our demonstration of strong spin-photon interaction is an important step towards coherent long-distance coupling of spin-qubits.

\end{abstract}

%\keywords{}

\maketitle

The ability to transmit quantum information over long distances is desirable for quantum information processors \cite{DiVincenzo2000}. Circuit quantum electrodynamics provides a well-established platform to connect distant qubits \cite{Majer2007}. There, microwave photons in a superconducting waveguide resonator couple to the electric dipole moment of multiple qubits fabricated in close proximity to the resonator. Strong coupling has been realized with superconducting qubits \cite{Wallraff2004} and, recently, with charge qubits in semiconductor quantum dots \cite{Bruhat2016, Mi2017a, Stockklauser2017}. However, charge qubits suffer from rapid decoherence due to charge noise \cite{Petersson2010, Hayashi2003}. A promising approach therefore takes advantage of the long coherence times of electron spins \cite{Hanson2007, Zwanenburg2013}. This approach comes with a major challenge as the coupling of photons to spins is several orders of magnitude weaker than the coupling to charge \cite{Schoelkopf2008}. The spin-photon interaction can be enhanced using materials with strong spin-orbit interaction \cite{Petersson2012}, devices with ferromagnetic leads \cite{Viennot2015}, or a magnetic field gradient generated by an on-chip micromagnet \cite{Pioro-Ladriere2008, Hu2012}. The use of exchange interaction to couple spin and charge is a different approach, realized in a three electron qubit \cite{Medford2013, Medford2013a, Gaudreau2011, Taylor2013, Russ2015}. Here, we implement such a three electron spin-qubit in a circuit quantum electrodynamics architecture \cite{Srinivasa2016, Russ2015a} hosted in GaAs and achieve strong spin-photon coupling as evident from the observation of vacuum Rabi mode splitting. Both the spin decoherence and the qubit-photon coupling strength can be controlled electrostatically \cite{Russ2016}.

Figure 1 shows optical (A) and scanning electron micrographs (B) of our hybrid quantum device. 
\begin{figure*}
\includegraphics[bb=0 0 309 241,width=0.75\linewidth]{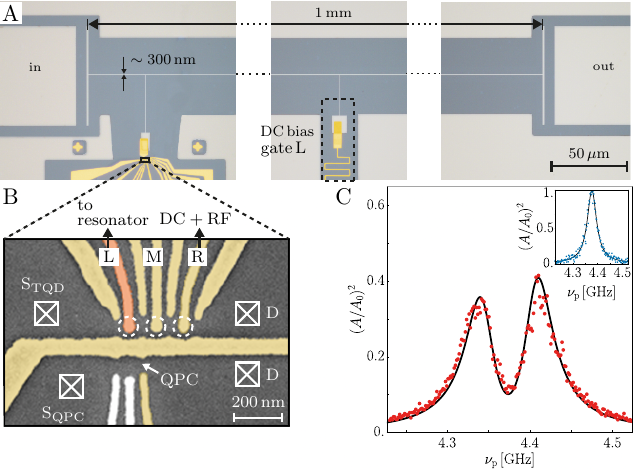}
\caption{Hybrid quantum device \& vacuum Rabi splitting. ({\bf A}) Optical micrograph of the device split into three pictures showing the resonator that is capacitively coupled to the input (in) and output (out) transmission lines. The region for DC bias that connects to the center of the resonator is indicated as a dashed black rectangle. ({\bf B}) False color scanning electron micrograph of the gate structure defined by electron beam lithography. The two white gates are kept at zero voltage in our experiments. The gate highlighted in orange is electrically connected to the resonator. The approximate positions of the left, middle and right quantum dots are indicated by dashed white circles. Their corresponding plunger gates are labelled as (L), (M) and (R). The right plunger gate is biased with DC and microwave signals. The triple quantum dot and quantum point contact have separate ohmic source contacts ($S_{\mathrm{TQD}}$ and $S_{\mathrm{QPC}}$) and a common drain contact (D). ({\bf C}) Resonator transmission $(A/A_0)^2$ as a function of resonator probe frequency $\nu_p$ for uncoupled (blue, inset) and coupled (red, main plot) configuration showing a strong spin-photon coupling vacuum Rabi mode splitting. The solid black lines are a fit to our input-output theory model.} 
\end{figure*}
Electrons are trapped in a triple quantum dot structure (TQD, see three dashed circles) by electrostatic confinement created by Au gates (Fig.~1B) on top of a GaAs/AlGaAs heterostructure. The heterostructure hosts a two dimensional electron gas (2DEG) in the TQD region $90\,\mathrm{nm}$ below the surface. The 2DEG has a mobility of $\mu=3.2\times 10^6\,\mathrm{cm^2/V s}$ and an electron density of $n_e=2.2\times 10^{11}\,\mathrm{cm^{-2}}$ at $4.2\,\mathrm{K}$. The electrostatic  potentials of the left, middle and right quantum dots are tuned with the respective plunger gate voltages $V_{L}$, $V_{M}$ and $V_{R}$. A quantum point contact (QPC) acts as a charge sensor that allows us to determine the TQD charge configuration. We operate the TQD as a three-electron spin-qubit \cite{Russ2015}, discussed in detail below.

To couple the qubit to microwave photons, the plunger gate of the left quantum dot extends to the superconducting microwave resonator shown in Fig.~1A. The left plunger gate is also DC-biased via a resistive Au line which is connected to the field anti-node of the center conductor of the resonator. The coupling strength $g$ between qubit and resonator photons is proportional to the square root of the characteristic impedance $\sqrt{Z_\mathrm{r}}$ of the resonator \cite{Devoret2007, Stockklauser2017}. It is enhanced by fabricating the resonator, shown in Fig.~1A, with a thin ($\sim 15\,\mathrm{nm}$) and narrow ($\sim 300\,\mathrm{nm}$) center conductor from the high kinetic-inductance material NbTiN \cite{Samkharadze2016}. We estimate $Z_\mathrm{r}=\sqrt{L_l/C_l}\sim 1.3\,\mathrm{k\Omega}$, with the resonator inductance (capacitance) $L_l\sim150\,\mathrm{\mu H/m}$ ($C_l\sim90\,\mathrm{pF/m}$) per unit length, resulting in a coupling strength enhancement by a factor of $5$ compared to a standard impedance-matched $Z_\mathrm{r}=50\,\mathrm{\Omega}$ resonator. Our choice of material and design allows us to operate the resonator in the presence of an external magnetic field applied parallel to the resonator plane \cite{Samkharadze2016}. In the experiments described here, we apply a magnetic field of $B_\mathrm{ext}=200\,\mathrm{mT}$.
\begin{figure*}
\includegraphics[bb=0 0 359 297,width=0.74\linewidth]{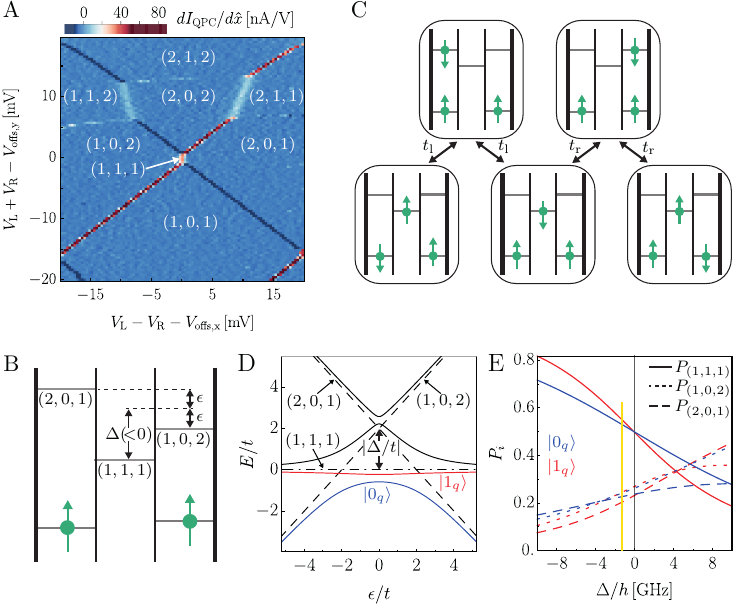}
\caption{TQD charge stability diagram \& spin-qubit operation regime. 
({\bf A}) Differential QPC current as a function of combinations of $V_\mathrm{L}$ and $V_\mathrm{R}$. $V_{\mathrm{offs,x}}$ and $V_{\mathrm{offs,y}}$ are voltage offsets, $\hat{x}$ is the quantity on the $x$-axis. ({\bf B}) TQD schematic defining the detuning parameters $\epsilon$ and $\Delta$. The three gray lines indicate the possible energy levels for the addition of the third electron. ({\bf C}) Illustration of relevant three electron states in the TQD that form the spin-qubit. The states mix via tunnel couplings $t_\mathrm{l}$ and $t_\mathrm{r}$. ({\bf D}) Eigenenergies of the system in ({\bf C}) as a function of $\epsilon/t$ for $\Delta/t=-2$ and symmetric tunnel couplings $t_\mathrm{l}=t_\mathrm{r}=t$. Dashed lines indicate the energy of the charge states (2,0,1) and (1,0,2) for $t_\mathrm{l}=t_\mathrm{r}=0$. The dash-dotted line is the eigenenergy of the $S=3/2$, $S_z=1/2$ state, which does not couple to any of the other states. The line also corresponds to the energy of the (1,1,1) states for $t_\mathrm{l}=t_\mathrm{r}=0$. The spin-qubit states $\ket{0_\mathrm{q}}$ (blue) and $\ket{1_\mathrm{q}}$ (red) are highlighted.
({\bf E}) Probabilities $P_{(1,1,1)}$ (solid lines), $P_{(2,0,1)}$ (dashed lines) and $P_{(1,0,2)}$ (dotted lines) as defined in the main text for $\ket{0_q}$ (blue) and $\ket{1_q}$ (red) as a function of $\Delta/h$. The plot is obtained for $t_\mathrm{l}/h=9.04\,\mathrm{GHz}$, $t_\mathrm{r}/h=7.99\,\mathrm{GHz}$ and $\epsilon/h=-1.03\,\mathrm{GHz}$. The position in $\Delta/h$ where Fig.~1C was recorded is highlighted (yellow line).
}
\end{figure*}

To demonstrate strong coupling of the spin-qubit with a microwave photon, we first detune the qubit transition frequency from the resonator resonance. We will explain below how the qubit energy can be tuned electrostatically. In this detuned situation, we determine the resonator frequency $\nu_\mathrm{r}=4.38\,\mathrm{GHz}$ and linewidth $\kappa/2\pi=47.1\,\mathrm{MHz}$ at an average resonator photon occupation of less than one (see inset of Fig.~1C). The photon population is determined using ac Stark shift measurements described later. When the spin-qubit is tuned into resonance with the resonator, we observe two distinct peaks in the transmission spectrum in Fig.~1C. The splitting of the resonator resonance into two well separated peaks, known as the vacuum Rabi mode splitting, is the characteristic fingerprint of strong coherent hybridization of a single microwave photon in the resonator and the spin-qubit in the TQD. From a fit of the vacuum Rabi splitting to an input-output theory \cite{Collett1984}, we extract the qubit-photon coupling strength $g/2\pi=(31.4\pm0.3)\,\mathrm{MHz}$ and the qubit decoherence rate $\gamma_2/2\pi=(19.6\pm0.5)\,\mathrm{MHz}$. 
As a result, our quantum device operates in the strong coupling regime, which is supported by the fact that the approximate peak separation is larger than the peaks width, i.e.  $2g > \kappa/2 + \gamma_2$. This is the main result of this article and more details on how it was achieved are described below.

The spin qubit is formed by tuning the TQD into the three electron regime. Figure 2A shows the charge stability diagram of the TQD, as measured by the charge detector. Regions of charge configurations $(k,l,m)$ are indicated, where the integers $k,l,m$ express the number of electrons in the three dots. The qubit operation point is located in the narrow (1,1,1) region between the (2,0,1) and (1,0,2) charge configurations. 
As illustrated in Fig.~2B, we introduce the asymmetry parameter $\epsilon$ and the detuning parameter $\Delta$ to quantify differences in the energies $E(i)$ of the three relevant charge configurations $i$ in the absence of inter-dot tunneling: $\epsilon=(E(2,0,1)-E(1,0,2))/2$ and $\Delta=E(1,1,1)-(E(2,0,1)+E(1,0,2))/2$.
Both parameters are tuned experimentally by plunger gate voltages: $\epsilon$ increases by increasing $V_\mathrm{L}$ and decreasing $V_\mathrm{R}$, while $\Delta$ increases by increasing $V_\mathrm{L}$ and $V_\mathrm{R}$ while decreasing $V_\mathrm{M}$.
Other charge configurations are not relevant, because the quantum dot charging energies are of the order of $1\,\mathrm{meV}$ ($240\,\mathrm{GHz}$), much larger than $k_\mathrm{B} T=3\,\mathrm{\mu eV}$ ($620\,\mathrm{MHz}$) for our experiments performed at an electronic temperature of $30\,\mathrm{mK}$.

In general, there are eight different spin configurations for three spins. For the asymmetric charge configurations (2,0,1) and (1,0,2), the three triplet states within the doubly occupied dots do not play a role because the singlet triplet splitting on the order of $1\,\mathrm{meV}$ ($240\,\mathrm{GHz}$) is much larger than temperature \cite{Hanson2007}. This leaves us with two relevant spin configurations for each of the two asymmetric charge configurations. Two of them, with $z$-component $S_z=1/2$ of total spin, are depicted in the top row of Fig.~2C. The other two are obtained by flipping the spin in the singly occupied dot giving $S_z=-1/2$. These spin configurations of the asymmetric charge configurations couple by tunneling to spin configurations of the (1,1,1) charge configurations as illustrated in Fig.~2C. Only the displayed spin configurations are relevant, because tunneling conserves both the total spin and its z-component. The configuration with $S_z=3/2$ is close in energy, but does not coherently couple to any other states and is therefore neglected here. Note that it also does not form the ground state of the system as the Zeeman energy of $5\,\mathrm{\mu eV}$ ($1.15\,\mathrm{GHz}$) is much smaller than the tunnel coupling of $\sim 33\,\mathrm{\mu eV}$ ($8\,\mathrm{GHz}$).

The qubit states are formed by a coherent superposition of the five basis states with $S_z=1/2$ depicted in Fig.~2C \cite{Russ2015}. An equivalent set of basis states with $S_z=-1/2$ differing just in the Zeeman energy exists, but is not depicted. Mixing between these different $S_z$-states by the Overhauser field of $\sim 5\,\mathrm{mT}$ \cite{Hanson2007} is suppressed by the much larger externally applied magnetic field.
The (1,1,1) states couple by exchange interaction between electrons in neighboring dots: an electron in the middle dot can be exchanged with an electron with opposite spin in the left (right) dot via tunneling to the asymmetric charge state.

The two lowest energy eigenstates of the system define the ground $\ket{0_\mathrm{q}}$ (blue) and the excited state $\ket{1_\mathrm{q}}$ (red) of the qubit with energy $E_\mathrm{q}(\Delta,\epsilon,t_\mathrm{l},t_\mathrm{r})=E_{\ket{1_\mathrm{q}}}-E_{\ket{0_\mathrm{q}}}$ (see Fig.~2D). For each of the qubit states $\ket{0_\mathrm{q}}$ and $\ket{1_\mathrm{q}}$, we define $P_{(1,1,1)}$ to be the sum of the occupation probabilities of the three (1,1,1) basis states, while $P_{(2,0,1)}$ and $P_{(1,0,2)}$ are the occupation probability of the (2,0,1) and (1,0,2) state, respectively. These quantities depend on $\Delta$ as depicted in Fig.~2E. Note that in Fig.~2E, $t_\mathrm{l}$, $t_\mathrm{r}$ and $\epsilon$ are the same as for the measurement of the vacuum Rabi mode splitting in Fig.~1C. The value of $\Delta/h=-1.44\,\mathrm{GHz}$ for the vacuum Rabi measurement is indicated in Fig.~2E with a vertical line, emphasizing the dominance of the spin-character of the qubit.
For $\Delta/h< 0\,\mathrm{GHz}$, where $P_{(1,1,1)}>0.5$, the qubit states $\ket{0_\mathrm{q}}$ and $\ket{1_\mathrm{q}}$ are dominated by the $(1,1,1)$ states, which can be calculated to be $\ket{0}=1/\sqrt{2}(\ket{\uparrow,\uparrow,\downarrow}-\ket{\downarrow,\uparrow,\uparrow})$ and $\ket{1}=1/\sqrt{6}(2\ket{\uparrow,\downarrow,\uparrow}-\ket{\uparrow,\uparrow,\downarrow}-\ket{\downarrow,\uparrow,\uparrow})$ [\hspace{1sp}\cite{Russ2015}, see Supplementary Information S1]. Quantum information is encoded into states that have equal charge but different spin configuration. Hence, this system corresponds to a spin-qubit \cite{Russ2017}. 

\begin{figure*}
\includegraphics[bb=0 0 357 378,width=0.76\linewidth]{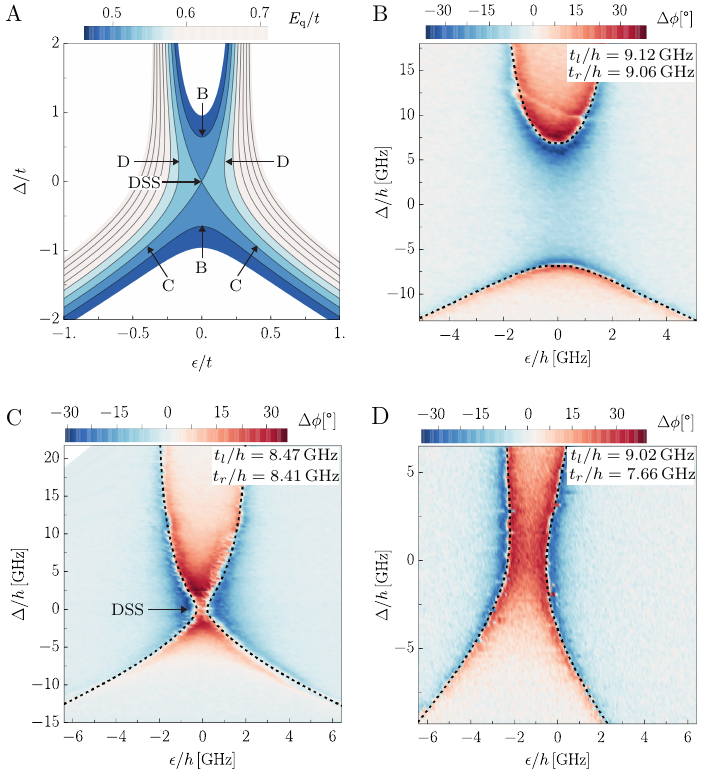}
\caption{Resonator response. ({\bf A}) Contour plot of the normalized qubit energy $E_\mathrm{q}/t$ for symmetric tunnel coupling $t$ as a function of $\epsilon/t$ and $\Delta/t$. The energy contours as probed in ({\bf B}-{\bf D}) are labelled. DSS marks the position of the double sweet spot. ({\bf B}-{\bf D}) Resonator phase response measured for $\nu_\mathrm{p}=\nu_\mathrm{r}$ for different tunnel coupling configurations. The dashed lines indicate a fit to the theory model.}
\end{figure*}

The finite qubit energy is a result of the exchange interaction of $(1,1,1)$ states mediated by the asymmetric charge states. Fig.~2E also shows that the admixture of the (2,0,1) and (1,0,2) charge configurations to the qubit states is tunable: for negative values of $\Delta$, the qubit is of spin character and dominated by the (1,1,1) charge configuration. However, for large positive values of $\Delta$, the qubit states are dominated by the asymmetric charge configurations (c.f. Fig.~2D) and are therefore of charge character. This admixture of asymmetric charge states enables coupling to the resonator electric field via electric dipole interaction \cite{Srinivasa2016, Russ2015a}.

To observe resonant qubit-photon interaction, we tune the spin-qubit energy by changing the detuning $\Delta$ and the asymmetry $\epsilon$ as shown in Fig.~3. 
We observe a phase shift of the microwave tone transmitted through the resonator whenever the qubit and the resonator approach a resonance $E_\mathrm{q}=h\nu_\mathrm{r}$. When the resonance is crossed, the phase changes sign. Experimentally, determining these transition points in the $\epsilon-\Delta$ plane at fixed tunnel couplings maps the energy contour $E_\mathrm{q}(\Delta,\epsilon)=h\nu_\mathrm{r}$, reproducing one of the theoretically expected energy contours shown in Fig.~3A. We can map different energy contour lines by changing the tunnel coupling. This is realized experimentally by changing the electrical potential of the gate lines between the plunger gates (see Fig.~1B). At constant and equal tunnel couplings, the qubit energy exhibits a saddle point at $\epsilon=\Delta=0$, called the double sweet spot (labelled DSS in the figure).
At this point, the qubit energy is insensitive to dephasing in $\epsilon$- and $\Delta$-directions to first order \cite{Russ2016}.

Through Figs.~3B-D, we increase the average tunnel coupling to map different contour lines of $E_\mathrm{q}$ as labeled in Fig.~3A. We obtain the magnitude of both tunnel barriers for Figs.~3B-D with a fit to the resonance positions of the phase response data. A simultaneous fit to the three datasets in Fig.~3B-D reduces the number of free parameters (see Supplementary Information S3) and results in excellent agreement between theoretical and measured resonance conditions. The qubit tunability allows us to observe qubit-photon interaction at the double sweet spot in Fig.~3C. Note that, as observed in Fig.~3D, the DSS is shifted for asymmetric barriers \cite{Russ2015}.

To further characterize the resonator-qubit interaction, we tune the qubit to a similar configuration as in Fig.~3C and record the resonator transmission spectra around the DSS. The transmission spectra as a function of $\Delta$ in Fig.~4A and $\epsilon$ in Fig.~4B show a clear anti-crossing of qubit and resonator. 
\begin{figure*}
\includegraphics[bb=0 0 332 249,width=0.72\linewidth]{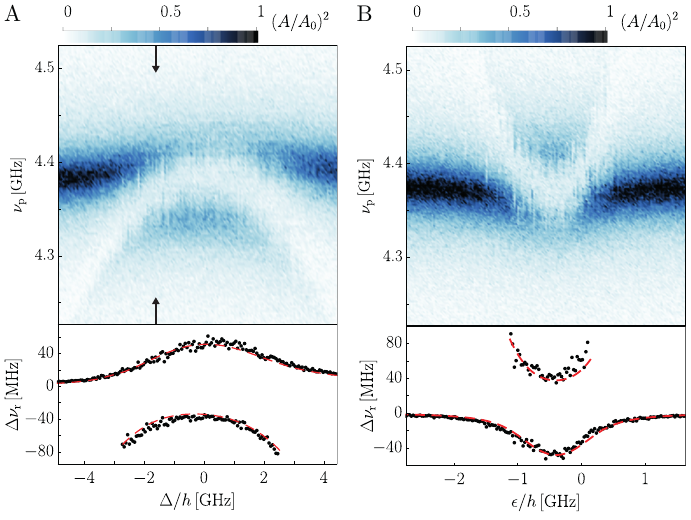}
\caption{Resonator spectrum at the DSS. ({\bf A}) Upper panel: resonator transmission as a function of probe frequency $\nu_\mathrm{p}$ and detuning $\Delta$ for $\epsilon/h=-1.03\,\mathrm{GHz}$, $t_\mathrm{l}/h=9.04\,\mathrm{GHz}$ and $t_\mathrm{r}/h=7.99\,\mathrm{GHz}$. The arrows indicate the position of the vacuum Rabi mode splitting as shown in Fig.~1C. Lower panel: frequency shift $\Delta\nu_\mathrm{r}$ of resonator resonance positions relative to the bare resonator frequency $\nu_\mathrm{r}$. The experimental points (dots) were extracted from a double Lorentzian fit to the data and are shown in comparison to theory (dashed line). ({\bf B}) Upper panel: resonator transmission as a function of detuning $\epsilon$ for $\Delta/h=0.23\,\mathrm{GHz}$, $t_\mathrm{l}/h=8.25\,\mathrm{GHz}$ and $t_\mathrm{r}/h=8.64\,\mathrm{GHz}$. The lower panel shows a comparison of experimental and theoretical resonance positions.}
\end{figure*}
The strong coherent hybridization of the spin-qubit and a single microwave photon in the resonator is also evident in the fully resolvable vacuum Rabi splitting over a large range of detuning $\Delta$ and asymmetry $\epsilon$. The vacuum Rabi splitting shown in Fig.~1C and discussed above is obtained for $\epsilon/h=-1.03\,\mathrm{GHz}$ and $\Delta/h=-1.44\,\mathrm{GHz}$ as indicated by the two arrows in Fig.~4A. The asymmetry of the spectrum in $\Delta$ with respect to the position of the maximum frequency shift of the higher frequency resonance in Fig. 4A reflects the $\Delta$-dependence of the coupling strength (discussed below). The shift of the extremum in Fig. 4A (4B) with respect to $\Delta=0$ ($\epsilon=0$) is due to the asymmetric tunnel coupling.
The transmission spectra also show that the DSS of the qubit is a saddle point in energy: in Fig.~4A we observe an energy maximum of the qubit around $\Delta\simeq 0$, in Fig.~4B the qubit energy has a minimum around $\epsilon\simeq 0$. In the lower panels of Fig.~4A and 4B, we show the measured peak positions in the transmission spectra with respect to the resonator frequency. The experimentally observed frequency shifts for both branches are in good agreement with our model as indicated by the dashed lines.

To further characterize the spin-qubit, we now consider the shift of the resonator frequency due to resonator-qubit coupling in the dispersive regime, where the qubit-resonator detuning is much larger than the qubit-photon coupling strength \cite{Schuster2005}. 
\begin{figure*}
\includegraphics[bb=0 0 329 180,width=0.78\linewidth]{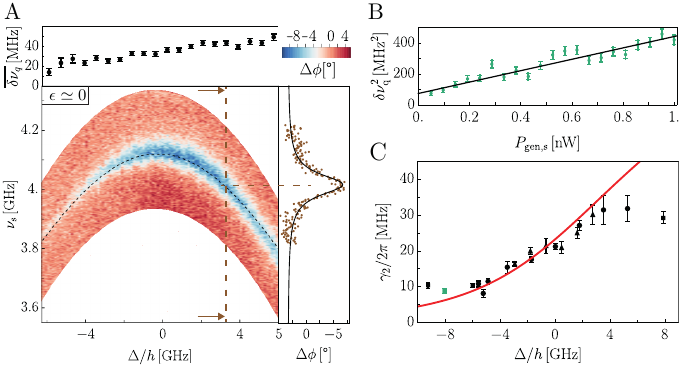}
\caption{Qubit spectroscopy. ({\bf A}) Phase response of the resonator probed at $\nu_\mathrm{p}=\nu_\mathrm{r}$ as a function of $\nu_\mathrm{s}$ and $\Delta$ around the DSS for $t_\mathrm{l}/h=8.10\,\mathrm{GHz}$, $t_\mathrm{r}/h=7.86\,\mathrm{GHz}$, a drive generator power of $P_{\mathrm{gen,s}}=0.75\,\mathrm{nW}$ and a resonator photon occupation of less than one. The theoretically expected position of the phase response minima is indicated by a dashed line. In the panel on the right, a Lorentzian with a HWHM ${\delta\nu_\mathrm{q}}$ (black line) is fit to a cut of the phase response (brown points). The panel on the top shows $\overline{\delta\nu_\mathrm{q}}$, which is the average of $\delta\nu_\mathrm{q}$ over five subsequent cuts along $\Delta$. ({\bf B}) Dependence of $\delta\nu_\mathrm{q}^2$ on drive generator power $P_{\mathrm{gen,s}}$ measured at $\Delta/h=-8.03\,\mathrm{GHz}$ and at the single sweet spot in $\epsilon$ for $t_\mathrm{l}/h=8.74\,\mathrm{GHz}$ and $t_\mathrm{r}/h=8.12\,\mathrm{GHz}$. ({\bf C}) Extracted qubit decoherence $\gamma_2/2\pi$ as a function of $\Delta$ for three different tunnel coupling configurations ($\blacksquare$ $t_\mathrm{l}/h=8.74\,\mathrm{GHz}$, $t_\mathrm{r}/h=8.12\,\mathrm{GHz}$, $\blacktriangle$ $t_\mathrm{l}/h=7.47\,\mathrm{GHz}$, $t_\mathrm{r}/h=7.77\,\mathrm{GHz}$, $\bullet$ $t_\mathrm{l}/h=8.10\,\mathrm{GHz}$, $t_\mathrm{r}/h=7.86\,\mathrm{GHz}$ in comparison to the theoretical prediction (line) for $t_\mathrm{l}/h=8.10\,\mathrm{GHz}$ and $t_\mathrm{r}/h=7.86\,\mathrm{GHz}$. Note that the theory prediction depends on tunnel coupling. The theory curves for the three tunnel coupling configurations almost overlap (see Supplementary Information Section S3), therefore only one curve is indicated in the plot. $\gamma_2$ obtained from the linear fit in ({\bf B}) is shown in green.}
\end{figure*}
In addition to the resonator probe tone at frequency $\nu_\mathrm{p}=\nu_\mathrm{r}$, a spectroscopy tone at frequency $\nu_\mathrm{s}$ is applied to the right plunger gate, indicated in Fig.~1B. At resonance with the qubit ($E_\mathrm{q}=h \nu_\mathrm{s}$), the drive excites the qubit from its ground state $\ket{0_\mathrm{q}}$ to the excited state $\ket{1_\mathrm{q}}$. This results in a dispersive shift of the resonator frequency that we detect as a drop in the phase response signal. By sweeping both the detuning $\Delta$ and the spectroscopy frequency $\nu_\mathrm{s}$, we trace the spectroscopic qubit signal in Fig.~5A. It resembles the $\Delta$-dependence of the qubit energy observed in Fig.~4A and calculated in Fig.~3A, and shows good agreement with theory (dashed line). Note that for the measurement in Fig.~4A $\epsilon$ is set to the energy minimum of the qubit (single sweet spot).

The qubit decoherence $\gamma_2/2\pi$ is equal to the half-width at half maximum (HWHM) $\delta\nu_q$ of the spectroscopic dip in the phase signal (right panel of Fig.~5A) in the limit of zero drive power ($P_{\mathrm{gen,s}}\rightarrow 0$) \cite{Schuster2005}. For finite drive power, such as in Fig.~5A, the spectroscopic signal is broadened \cite{Schuster2005}. We define the HWHM $\overline{\delta \nu_q}$ as the average of $\delta\nu_q$ for five subsequent cuts along the $\Delta$-direction in Fig.~5A, as illustrated in the top panel in Fig.~5A. Its increase can therefore be interpreted as an enhanced qubit decoherence arising when we increase the admixture of the asymmetric charge states $(2,0,1)$ and $(1,0,2)$ (see Fig.~2E). 

We extract $\gamma_2$ by measuring $\delta\nu_q$ as a function of the power of the spectroscopy tone (see Fig.~5B) for different $\Delta$ and three different sets of coupling configurations. The results plotted in Fig.~5C are consistent with the spectroscopic measurement in Fig.~5A where we observe an increase of $\gamma_2$ with $\Delta$. For a high admixture of asymmetric charge states, we measure a maximum decoherence rate $\gamma_2/2\pi\sim 30\,\mathrm{MHz}$. For a more (1,1,1)-like character of the spin-qubit, we extract a minimum decoherence rate $\gamma_2/2\pi\simeq 10\,\mathrm{MHz}$, which corresponds to a dephasing time $T_2^\star=1/\gamma_2=16\,\mathrm{ns}$.

The theoretical value of $\gamma_2$ is obtained considering the effects of charge noise originating from voltage fluctuations. For values of $\Delta/h\sim 0\,\mathrm{GHz}$, noise due to voltage fluctuations strongly mixes the qubit states with the asymmetric charge states, leading to decoherence. For more details on the model we refer to the Supplementary Information S1. The influence of the asymmetric charge states on decoherence is well described by our charge noise model, apart from the limits of large positive or negative $\Delta$. A possible explanation for the lower limit of $\gamma_2$ is inhomogeneous broadening due to coupling of the spin-qubit to the hyperfine field in the GaAs host material, consistent with previous work that reported a similar dephasing time for a resonant exchange qubit \cite{Malinowski2017}. The discrepancy for large positive $\Delta$ might be due to phonon induced qubit relaxation \cite{Mehl2013}.

Finally, we show that the average photon number in the resonator is well below one for the measurement of the Rabi splitting. In the dispersive regime the qubit frequency $\nu_q$ shifts as a function of the number of photons $n$ in the resonator, which linearly depends on the resonator generator power $P_{\mathrm{gen,r}}$. 
\begin{figure*}
\includegraphics[bb=0 0 314 171,width=0.75\linewidth]{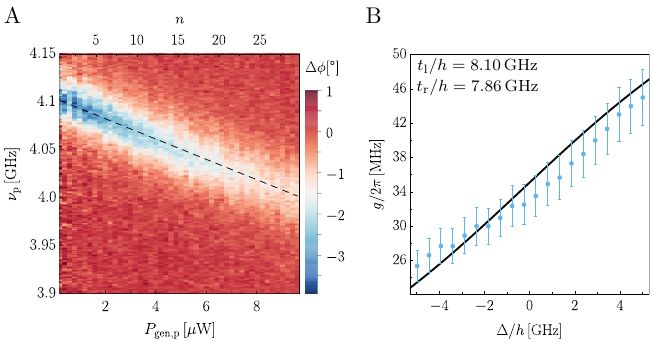}
\caption{ac Stark shift. ({\bf A}) Phase response as a function of spectroscopy frequency $\nu_\mathrm{s}$ and resonator probe generator power $P_\mathrm{gen,p}$. The resonator power is converted to the average number of photons in the resonator $n$. The generator of the drive gate is set to a power of $P_\mathrm{gen,s}=0.25\,\mathrm{nW}$, the resonator is probed on resonance ($\nu_\mathrm{p}=\nu_\mathrm{r}$). The qubit parameters are $t_\mathrm{l}/h=8.72\,\mathrm{GHz}$, $t_\mathrm{r}/h=8.18\,\mathrm{GHz}$, $\Delta/h=-6.0\,\mathrm{GHz}$ and $\epsilon/h=-0.26\,\mathrm{GHz}$. The position of the phase response minima are indicated with a dashed line. ({\bf B}) Spin-qubit photon coupling strength $g$ as a function of $\Delta$ (points) compared to the theory prediction (line) for $\epsilon$ close to the single sweet spot.}
\end{figure*}
In addition, there is a Lamb shift of the qubit frequency due to the coupling to vacuum fluctuations. This results in the dressed qubit frequency $\tilde\nu_q=\nu_q+(2n+1)(g/2\pi)^2/(\nu_q-\nu_\mathrm{r})$ \cite{Schuster2005}. In Fig.~6B, we observe the frequency shift due to the ac Stark shift in the spectroscopic qubit signal measured at $\Delta/h=-6.02\,\mathrm{GHz}$ and $\epsilon/h=-0.26\,\mathrm{GHz}$. At this operating point, we obtain $g$ from an independent resonator frequency shift measurement similar to the one displayed in  Fig.~4B (see Supplementary Information S4). From a linear fit to the power dependent dressed qubit frequency in Fig.~6B, we obtain the calibration factor $\alpha\equiv n/P_{\mathrm{gen,r}}\simeq3\times 10^{-3}\,\mathrm{photons/nW}$. The vacuum Rabi splitting shown in Fig.~1C was recorded for $P_{\mathrm{gen,r}}=100\,\mathrm{nW}$. We can therefore reliably claim that for this measurement the average number of photons in the resonator is on the order of $\sim 0.3$. This confirms that we indeed achieved a strong hybridization of the spin-qubit with a single microwave photon.

With the known calibration factor $\alpha$, the ac Stark shift gives direct access to the qubit-photon coupling strength (see Supplementary Information S3).
%which expresses as $g=2\pi\sqrt{a(\nu_q-\nu_\mathrm{r})/2 \alpha}$ using the slope $a$ of its power dependence. Here, the bare qubit frequency $\nu_q$ is obtained by interpolating the dressed qubit frequency $\tilde\nu_q$ to zero generator power. 
We observe in Fig.~6B, that the coupling strength increases with increasing $\Delta$. As the contribution of (1,0,2) and (2,0,1) charge configurations to the qubit states increases with $\Delta$, the electric dipole moment of the qubit states and hence the qubit-resonator coupling is enhanced. This increase in coupling strength, however, comes at the cost of an increase in qubit decoherence (see Fig.~5C). Our theoretical model describes this behavior quantitatively.

We have coherently coupled a TQD spin-qubit to single microwave photons in a circuit quantum electrodynamics architecture. The TQD spin-qubit arises from exchange interaction which couples spin and charge independent of the host material. Other spin-qubit implementations are restricted to materials with strong spin-orbit interaction \cite{Petersson2012} or require additional components such as ferromagnets \cite{Viennot2015,Mi2017} for the spin-charge hybridization. Furthermore, the TQD spin qubit is versatile as all its parameters can be controlled electrostatically.
For these reasons, it is possible to move our architecture to material systems with minimal hyperfine interaction, such as graphene \cite{Trauzettel2007} or isotopically purified silicon \cite{Zwanenburg2013} without the necessity to deposit ferromagnetic materials, which is generally undesirable in the presence of a superconductor.
By doing so, we expect the qubit coherence to improve by at least one order of magnitude.\\\\
While writing up our results we became aware of independent but related work by another group demonstrating strong spin-photon coupling in a double quantum dot spin-qubit in silicon \cite{Mi2017}.

\begin{acknowledgments}
We acknowledge helpful discussions with Maximilian Russ and Anna Stockklauser. We would also like to thank Michele Collodo, Philipp Kurpiers and Peter M\"arki for valuable contributions to our experimental setup. This work was supported by the Swiss National Science Foundation through the National Center of Competence in Research (NCCR) Quantum Science and Technology. U.C.M. and A.B. were supported by NSERC and the Canada First Research Excellence fund.
\end{acknowledgments}

\bibliographystyle{apsrev4-1}
\bibliography{Paper_references_V3}

\end{document}